\documentclass[12pt]{iopart}
\usepackage{times}  
\usepackage{graphicx}

\begin{document}

\title[Environment-dependent potential for solid Ar]{A simple
environment-dependent overlap potential and Cauchy violation in solid argon}

\author{Masato Aoki and Tatsuya Kurokawa}
\address{Faculty of Engineering, Gifu University
Yanagido, Gifu 501-1193, Japan}
\ead{masato@gifu-u.ac.jp}

\begin{abstract}
We develop an analytic and environment-dependent
interatomic potential for the overlap repulsion in
solid argon, based on an approximate treatment
of the non-orthogonal Tight-Binding theory for the 
closed-shell systems. 
The present model can well
reproduce the observed elastic properties of solid argon
including
Cauchy violation at high pressures, yet very simple.
A useful and novel analysis
is given to show how the elastic properties 
are related to the environment-dependence incorporated
into a generic pairwise potential. 
The present study has a close link to 
the broad field of computational materials science,
in which the inclusion of  environment dependence 
in short-ranged repulsive part of
a potential model is sometimes crucial in
predicting the elastic properties correctly.
\end{abstract}

%Uncomment for PACS numbers title message
% Keywords required only for MST, PB, PMB, PM, JOA, JOB? 
%\vspace{2pc}
%\noindent{\it Keywords}: Article preparation, IOP journals
% Uncomment for Submitted to journal title message
%\submitto{\JPA}
% Comment out if separate title page not required
%\maketitle

\section{Introduction
\label{introduction}}

%%% experimental facts
Recent progress of Brillouin spectroscopy 
at very high pressures\cite{ShimizuTKS01,Grimsditch-etal86}
has revealed that interatomic forces
in  fcc solid argon must be far beyond any kinds of two-body, 
central force model. 
Shimizu {\it et al.}\cite{ShimizuTKS01}  precisely measured 
a large violation of the Cauchy 
relation up to 70 GPa and stressed the important role of 
many-body forces, in order to construct good potentials for
high-density noble gases, which should be crucial in understanding
their behavior in planetary bodies by means of molecular simulations. 

%%% Cauchy relation and many-atom forces
The Cauchy relation\cite{BornHuang88} for the elastic constants
of cubic crystals at a hydrostatic pressure $P$ is given by
$C_{12} - C_{44} -2P =0$,
which must be satisfied in centrosymmetric cubic crystals, 
including the fcc solid argon, if the total energy
is given by sum of purely pairwise terms.
The deviation from it is therefore a measure of the many-atom nature 
of interatomic interactions. The Aziz-Slaman 
model for high pressure argon\cite{Aziz-Slaman90}, which might be one of the 
most sophisticated yet simple ones thus far proposed,
fails to reproduce any violation of Cauchy relation, 
simply because the model is pairwise.

%%% ab initio approach is okey, but we need simpler potentials
On the other hand, the {\it ab initio} Density Functional Theory (DFT) approach 
using the pseudopotential planewave method\cite{Iitaka-E-02}, and that with 
projector-augmented wave implementation for core electrons\cite{Tse-etal-02},
and linear muffin-tin orbital (LMTO) method\cite{Tsuchiya-K-02,Aoki03}
have successfully reproduced the observed  elastic constants,
as well as the density of the solid argon over the measured 
range of the pressure. From these theoretical results, it should be 
a natural and logical consequence that one would expect
to have an simple and reasonably accurate model for the many-atom
interaction in condensed argon, by coarse-graining the {\it ab initio} 
electronic models to a rather empirical atomistic model.

%%% why many-atom ?
The importance of many-body forces in solid argon at 
high pressures has been examined by several authors 
along the idea of many-atom 
expansion\cite{Rosciszewski-etal00,Lotrich-S97,Schwerdtfeger-etal06}, 
in which one
assumes that the total energy is `additively' decomposable
into $N$-atom ($N=2,3,4\cdots$) terms plus the additional
energy of zero-point vibrations, and that 
the expansion is well convergent. The three-atom contribution 
from the exchange energy was emphasized\cite{Lotrich-S97}
because it stabilise argon in  fcc structure, rather than the hcp,
which is predicted by all the available pairwise models without 
the zero-point energy\cite{Schwerdtfeger-etal06}. 
However, it is pointed out that the convergence of
this type of expansion becomes worse in a situation 
in which many-atom effect is more important\cite{Schwerdtfeger-etal06}.

\begin{figure}[tb]
\begin{center}
\includegraphics[scale=0.65, trim=0 30 0 0]{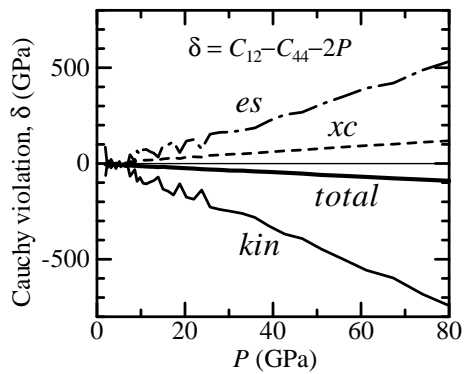}
\end{center}
\caption{Total Cauchy violation $\delta$, decomposed into contributions 
from the kinetic (thin solid), electrostatic (dot-dashed), 
and exchange-correlation (dashed) energies.}
\label{ab-initio-C-decomp}
\end{figure}

%%% learning from the ab initio decomposition
Figure \ref{ab-initio-C-decomp} shows 
the Cauchy violation defined by
\begin{equation}
\delta \equiv C_{12} - C_{44} -2P ,
\label{Cauchy-deviation}
\end{equation}
and its breakdown into the contributions 
from kinetic, electrostatic and exchange-correlation energies
predicted\cite{Aoki03}  by using all-electron calculation 
within DFT\cite{SavrasovS92}.
Each curve is plotted as a function of the total pressure.
Clearly,  the central role for the observed negative $\delta$ 
is played by kinetic energy,
which remains after the large cancellation by the opposite contributions
from electrostatic and exchange-correlation. 
It should be noted that non-zero contribution from the electrostatic
energy immediately excludes a primitive picture of overlapping frozen 
atomic charge-density. Therefore, it is implied that deformation of 
density (wavefunctions) should be relevant.

%%% purpose
The purpose of this paper is to develop 
an analytic interatomic potential for solid argon, that is based on 
the quantum mechanics of electrons, and that can well
reproduce the observed elastic properties including
the Cauchy violation at high pressures,
yet that is made as simple as possible.
The problem we are to treat now has a close link to 
the broad field of computational materials science,
since the inclusion of  environment dependence 
in short-ranged repulsion is sometimes 
crucial\cite{TangWCH96,HaasWFEH98,MrovecNPV04,DucPZV88}
in obtaining reliable and transferable models
for simulations in empirical and semi-empirical
approaches.

In section 2, a Tight-Binding description for overlap repulsion 
in closed-shell atoms,
which depends on atomic environment, will be presented as the
theoretical base that underpins more empirical and analytic model.
General properties of repulsive potential with
environment-dependent parameters are analysed, 
and a simple functional form for the overlap repulsion
is designed for argon  and proposed in section 3.
The result of the fitted analytic potential is presented in 
section 4, followed by a concluding section 5.

\section{Environment-dependent overlap repulsion: 
Tight-Binding description
\label{section2}}

A system of closed-shell atoms may be well described within 
the non-orthogonal Tight-Binding Bond Model 
(TBBM)\cite{Finnis03,SFPO88},
 in which  the total binding energy is given by
\begin{equation}
E_{\mathrm{B}}=E_{\mathrm{cov}} + E_{\mathrm{ren}}
+E_{\mathrm{rep}}+E_{\mathrm{vdW}} .
\label{EB-TBBM-vdW}
\end{equation}
The first term is the covalent energy
\begin{equation}
E_{\mathrm{cov}} = 
2\mathrm{Tr}[\mathsf{H}\mathsf{S}^{-1}]
-2\mathrm{Tr}[\mathsf{H}] 
=-2\mathrm{Tr}[\mathsf{H}\mathsf{O}\mathsf{S}^{-1}] ,
\label{Ecov-closed-shell}
\end{equation}
where $\mathsf{H}$ and $\mathsf{S}=\mathsf{1}+\mathsf{O}$ are the Hamiltonian 
and overlap matrices defined by
\begin{equation}
(\mathsf{H})_{i\mu,j\nu}%=\langle{\psi_{i\mu}}|\hat{H}|{\psi_{j\nu}}\rangle,\ \ \ \
=\int \psi_{i\mu}^{*}(\mathbf{r})\hat{H}\psi_{j\nu}(\mathbf{r})\mathrm{d}^3r
\label{H-matrix-elements}
\end{equation}
and
\begin{equation}
(\mathsf{S})_{i\mu,j\nu}%=\langle{\psi_{i\mu}}|{\psi_{j\nu}}\rangle
= \delta_{i,j}\delta_{\mu,\nu} + (\mathsf{O})_{i\mu,j\nu}
=\int \psi_{i\mu}^{*}(\mathbf{r})\psi_{j\nu}(\mathbf{r})\mathrm{d}^3r, \ \ \ \
\label{S-matrix-elements}
\end{equation}
in a basis set of atomic orbitals $\psi_{i\mu}$, where $\mu$ runs 
over orbitals on site $i$.
The spin degeneracy enters as the prefactor 2 before 
the usual symbol $\mathrm{Tr}$ for the trace. 
The Hamiltonian operator $\hat{H}$ refers to the density of
superposition of frozen atomic densities\cite{SFPO88,Finnis03}.
Note that the inverse overlap matrix, $\mathsf{S}^{-1}$, 
in Eq.(\ref{Ecov-closed-shell})
is equivalent to the density matrix  in this case
of fully occupied system.
The second term in Eq.~(\ref{EB-TBBM-vdW}), $E_{\mathrm{ren}}$, 
is defined by
\begin{equation}
E_{\mathrm{ren}} = 
2\mathrm{Tr}[\mathsf{H}]-2\mathrm{Tr}[\mathsf{H}_\mathrm{fa}],
\label{Eren}
\end{equation}
which accounts for the on-site energy shift due to the contraction 
or localisation of free atomic orbitals on going into a condensed 
environment, and $\mathsf{H}_\mathrm{fa}$ is a diagonal matrix of
the free atomic energy levels.
$E_{\mathrm{rep}}$ represents the contribution from
the change in the electrostatic and exchange-correlation energies
associated with {\it frozen} atomic charge densities as they are brought 
together from the free space. Thus, this term 
is environmentally {\it independent} by construction, and 
usually approximated as a sum of repulsive pair-wise 
potentials between atoms. 
$E_{\mathrm{vdW}}$, added supplementarily to TBBM, 
denotes the van der Waals potential, which may 
be approximated using the empirical pair-wise
inverse power function of interatomic separation, that is
$-c_6R_{ij}^{-6}$.

In order to illustrate that Eq.(\ref{Ecov-closed-shell}) 
essentially represents the overlap repulsion, we now consider only 
the outermost $p$-shell states as the basis,
and then the bond integral\cite{Pettifor95} part, $\mathsf{B}$, may be separated 
from $\mathsf{H}$ as
\begin{equation}
\mathsf{H}=\epsilon_{p}\mathsf{S}+\mathsf{B}, 
\end{equation}
where $\epsilon_{p}$ is the common diagonal element of $\mathsf{H}$.
Further simplification can be made exploiting 
the Wolfsberg-Helmholtz\cite{Pettifor95} (or extended H\"uckel) 
approximation to write $\mathsf{B}=-b\mathsf{O}$ using a constant $b>0$.
We easily find 
\begin{equation}
E_{\mathrm{cov}} 
= 2b\mathrm{Tr}[\mathsf{O}^2(\mathsf{1}+\mathsf{O})^{-1}],
\label{Ecov-overlap-rep}
\end{equation}
and see immediately that $E_{\mathrm{cov}}=0$ if the basis is orthogonal
({\it i.e.} $\mathsf{O}=$ zero), and it turns into
repulsive when the $\mathsf{O}$ matrix is switched on.
Note that Eq.(\ref{Ecov-overlap-rep}) has no explicit 
dependence on $\epsilon_{p}$.
The name of this term, therefore, is only nominal for the noble gases 
as it gives overlap repulsion rather than the covalent bonding.
The lowest order term in Eq.~(\ref{Ecov-overlap-rep}) is given by
$2b\mathrm{Tr}[\mathsf{O}^2]$, which may be broken down into contributions
in a purely pair-wise form
\begin{equation}
\Phi_{ij}(R_{ij}) = 4b\{|O_{pp\sigma}(R_{ij})|^2+2|O_{pp\pi}(R_{ij})|^2\},
\label{Ecov-overlap-rep2}
\end{equation}
where $O_{pp\sigma}(R_{ij})$ and $O_{pp\pi}(R_{ij})$ are $\sigma$- and 
$\pi$-overlap integrals along interatomic separation $\mathbf{R}_{ij}$.
The higher order correction terms, which  arise from multiplication
of $(\mathsf{1}+\mathsf{O})^{-1}$ in Eq.~(\ref{Ecov-overlap-rep}), 
or alternatively, multiplication
of $(\mathsf{1}+\mathsf{O})^{-1/2}$ from both sides of $\mathsf{O}^2$
in a symmetric L\"owdin form, would introduce many-atom effects as 
derived by Nguyen-Manh {\it et al.} for 
environment-dependent bond integrals\cite{DucPV00}.
For simplicity of our model, we may neglect these higher order corrections
 to write
\begin{equation}
E_{\mathrm{cov}} \cong \mathrm{Tr}[-2\mathsf{H}\mathsf{O}] 
= {\textstyle\frac{1}{2}}\sum_{i\ne j}\Phi_{ij}(R_{ij}) \equiv E_{\mathrm{ovl}}.
\label{Ecov-Phi}
\end{equation}

Assuming the Slater-type atomic $p$-orbitals with exponential tail of
$\exp({-\kappa_{i}r})$ for atom $i$, the bond and overlap integrals 
decay like 
$\exp[{-(\kappa_{i}+\kappa_{j})R_{ij}}]$ and the overlap potential 
Eq.~(\ref{Ecov-overlap-rep2}) takes form of
\begin{equation}
\Phi_{ij}(R_{ij})= 
(\mbox{polynomial of $R_{ij}$})\times
\exp[{-(\kappa_{i}+\kappa_{j})R_{ij}}]
\label{Phi(R)-from}
\end{equation}
An important environment effect can naturally be taken into account if 
we think of $\{\kappa_{i}\}$ as a set of variational parameters.
It is a straightforward exercise to show for a
hydrogen-like atom with an effective atomic number $Z^*$ that
(in atomic Rydberg units)
\begin{equation}
\int \psi_{i\mu}(\mathbf{r})(-\nabla^2-2Z^{*}/r)
\psi_{i\mu}(\mathbf{r})\mathrm{d}^3r
=(\kappa_i - Z^*/2)^2 -(Z^*/2)^2
\label{Hydrogen-like}
\end{equation}
with $-(Z^*/2)^2$ being the lowest $p$-energy level, 
we see that the parabolic 
penalty for an augmented $\kappa$ arises from increase 
in kinetic energy due to localisation. 
This parabolic behaviour occurs as a result of the change in the
effective radius of the atomic wavefunction, regardless of 
particular form of the atomic pseudopotentials.
Therefore, we may assume for the diagonal matrix elements of $\hat{H}$ that
\begin{equation}
(\mathsf{H})_{i\mu,i\mu}=(\kappa_i - \kappa_{i0})^2 + \epsilon_{i0}
=\epsilon_{ip}(\kappa_i),
\label{Hii}
\end{equation}
where $\kappa_{i0}$ and $\epsilon_{i0}$ are constants, and thus,
$E_{\mathrm{ren}}$ has the role of penalty for localisation
of atomic orbitals through Eq.~(\ref{Hii}), while the localisation
will reduce the magnitude of overlap repulsion.
Therefore the optimum values of $\{\kappa_{i}\}$ will be determined
by minimizing $E_{\mathrm{ovl}}+E_{\mathrm{ren}}$ with
respect to each $\kappa_{i}$. 
In the case of $p$-shells, given a simplified form for overlap potential
$\Phi_{ij}=6q\exp[{-(\kappa_{i}+\kappa_{j})R_{ij}}]$ with a
constant $q$, this minimisation leads to
\begin{equation}
\Delta\kappa_i = \kappa_i - \kappa_{i0}
= \sum_{j(\neq i)}qR_{ij}
\exp[{-(\kappa_{i}+\kappa_{j})R_{ij}}].
\label{kappa-eqn}
\end{equation}

The set of solutions is indeed environment-dependent and
we see that the constant $\kappa_{i0}$ is the solution 
for the limiting case of infinitely separated atoms.
Since the penalty due to increase in the kinetic energy is
steep, $\Delta\kappa_i/\kappa_{i0}$ would be small enough
to replace $\kappa_i$ in the exponential in Eq.(\ref{kappa-eqn})
with $\kappa_{i0}$. This solves the equation to give 
explicit $\Delta{}\kappa_i$ that is exact to first order. 
The energy increase $\Delta{}E_{\mathrm{ren}}$
due to this minimisation is given exactly by the sum of 3$(\Delta\kappa_i)^2$,
which partially sets off and just halves the lowest order decrease 
in overlap energy $\Delta{}E_{\mathrm{ovl}}$. The resultant lowering, 
$\frac{1}{2}\Delta{}E_{\mathrm{ovl}}$, could be obtained by employing 
$\frac{1}{2}\Delta\kappa_i$ instead of $\Delta\kappa_i$
in the overlap energy. This treatment {\it eliminates}
$\Delta{}E_{\mathrm{ren}}$ and simplifies the functional form of 
the potential, which is to be proposed in the next section.

The environmental effect that we are looking at 
is a tendency that more contracted atomic orbitals 
are preferred in a denser environment. The physics behind it
has been beautifully justified in the pioneering work by
Skinner and Pettifor\cite{Skinner-P-91},
who have implemented the chemical pseudopotential theory
using the orbital exponent as a variational parameter
within the Harris-Foulkes scheme\cite{Harris85,Foulkes89}, 
and found that
the orbital exponents ($\kappa$'s in our notation) 
of hydrogen atoms in molecule, simple cubic and fcc lattices 
are strongly environment-dependent as stated above.

\section{Analytic model for solid argon}

Let us first analyse some general properties of a
repulsive potential that is a function of
environment-dependent parameters as well as the distance.

Provided that the functional form of 
repulsive interatomic potential is given by
\begin{equation}
 \Phi_{ij}(R_{ij};\lambda_{i}+\lambda_{j})
\label{eq:Phi}
\end{equation}
with the environment-dependent parameters ($\lambda$'s) that 
are written as a sum of pairwise functions, namely
\begin{equation}
 \lambda_i=\sum_{k(\neq i)}\rho(R_{ik}).
\label{eq:lanmda_i}
\end{equation}
Cauchy violation can be calculated analytically  
for cubic lattice. The result is written\cite{Aoki03}
\begin{equation}
  \delta 
=
\frac{2}{9\Omega}
 \left[ \alpha_0^2\sum_{j(\neq 0)}
  \frac{\partial^2\Phi_{0j}}{\partial{}\lambda_0^2}
+ 
  \alpha_0\!\!\sum_{j(\neq 0)}
  R_{j}\frac{\partial^2\Phi_{0j}}{\partial{}R_{j}\partial{}\lambda_0}
 \right]
   \label{eq:delta-analytic} \\ 
\end{equation}
with 
\begin{equation}
\alpha_0 =  \sum_{k(\neq 0)}R_{k}\rho'(R_k),
\end{equation}
where $\Omega$ is the atomic volume and  
$R_{j}=R_{0j}=\sqrt{x_j^2 + y_j^2 + z_j^2}$ 
is the distance to atom $j$ from the central 
atom $i=0$ at the origin. The prime on $\rho$ denotes the distance derivative.
Note that all lattice sites are equivalent under 
a homogeneous strain.
$\alpha_0$ represents the strength of environmental effect on atom $0$.
Since $\rho(R)$ at this point is completely arbitrary, we may 
assume that it is a positive and monotonically decreasing function
of distance in the range of interest; hence $\alpha_0<0$. 
It may also be a physically reasonable assumption that
the repulsive potential $\Phi_{0j} (>0)$ is also a monotonically 
decreasing function
of distance in the range of interest.
We see from Eq.(\ref{eq:delta-analytic}) for the case of very weak 
$\alpha_0$ that negative Cauchy violation occurs
if $\partial^2\Phi_{0j}/\partial{}R_{j}\partial\lambda_0>0$.
This condition is likely to be fulfilled, since an environmental effect
tends to weaken the overlap repulsion to give 
$\partial\Phi_{0j}/\partial\lambda_0<0$, as we have seen 
in the previous section. 
The expression for the pressure from $\Phi$ is given by
\begin{equation}
P =  -\frac{1}{6\Omega}
 \left[ \sum_{j(\neq 0)}\!\!R_j
  \frac{\partial\Phi_{0j}}{\partial{}R_j}
  +2\alpha_0\!\!\sum_{j(\neq 0)}
  \frac{\partial\Phi_{0j}}{\partial\lambda_0}
 \right].
\end{equation}
The first term in the square bracket represents
the pairwise component.
We see environmental effect, the second term, causes 
reduction in pressure as expected.

A simple functional form of 
overlap repulsion that takes into account the physics of
the environment effect as we have discussed is now proposed, 
that is,
\begin{equation}
 \Phi_{ij}(R_{ij};\lambda_i+ \lambda_j)
 = \exp({-\lambda_i})\exp({-\lambda_j})V_{R}(R_{ij}),
\label{eq:separable-Phi}
\end{equation}
where $V_{R}$ is environmentally independent pairwise 
function. The environmental effect on site $i$ is
entering in a very simple separable form by a factor 
$\exp({-\lambda_i})$, which corresponds to the contraction 
factor $\exp({-\Delta\kappa_i}R_{ij})$. However,
the direct dependence on the particular length $R_{ij}$
 has been dropped for simplicity.
This manner of parameterisation for the environmental effect 
can also be seen in the `breathing-shell model' 
(See Ref. \cite{Sangster73,Matsui05} and references therein)
and `compressible ion model'\cite{Marks-FHP-01} for
oxides, such as MgO. The both models are provided with 
the parameters that correspond to the variation in 
effective size of ionic cores, and  reduction in 
core size causes exponential reduction as $\lambda_i$
in the present model does. It should be noted that
a kind of penalty function 
has been eliminated from the present model as it was justified
in the previous section.
The contraction factor of type $\exp({-\Delta\kappa_i}R_{ij})$
was modelled by Nguyen-Manh {\it et al.} in the form of screened Yukawa-type 
potential\cite{DucPZV88,MrovecNPV04} and it was used to 
explain Cauchy pressures in transition metals intermetallics
within a Tight-Binding and Harris-Foulkes approaches.

It follows from the functional form proposed above that
full expressions for the pressure $P$, Cauchy violation $\delta$,
adiabatic bulk modulus $B$,
and the cubic elastic constants $C_{11}, C_{12}, C_{44}$ are given by
\begin{eqnarray}
 P &=& {\textstyle\frac{1}{3}}(-v+2u\alpha_0),
 \label{eq:P} \\
 \delta &=& {\textstyle\frac{4}{9}}(-\alpha_0 v + u\alpha_0^2),
 \label{eq:delta} \\
 B &=& {\textstyle\frac{2}{3}}P 
  + {\textstyle\frac{1}{3}}K + \delta, 
  \label{eq:Bs} \\
 C_{11} &=& -P + P^s + K^s + \delta,
  \label{eq:C11PKd} \\
 C_{12} &=&  {\textstyle\frac{1}{2}}(3P +K -P^s-K^s) +\delta,
  \label{eq:C12PKd} \\
 C_{44} &=& {\textstyle\frac{1}{2}}(-P +K -P^s-K^s)
  \label{eq:C44PKd} 
\end{eqnarray}
with 
\begin{eqnarray}
 P^s &=& {\textstyle\frac{1}{3}}(-v^s+2u\alpha^s_0) ,
 \label{eq:Ps} \\
 K &=& {\textstyle\frac{1}{3}}(w-2u\beta_0), \ \ \  
 K^s = {\textstyle\frac{1}{3}}(w^s-2u\beta^s_0) ,
 \label{eq:KKs} 
\end{eqnarray}
where $u$ equals the energy density and $v,w$ also are 
similar quantities determined by first- and second-order derivatives 
of the potential, and $u^s,v^s,w^s$ are weighted sums.
These are defined by
\begin{eqnarray}
 &&u = \frac{1}{2\Omega}\sum_{j(\neq 0)}\Phi_{0j},\ \ \ \ \
 u^s = \frac{1}{2\Omega}\sum_{j(\neq 0)}\Phi_{0j}s_j,
 \label{eq:uus} \\
&& v = \frac{1}{2\Omega}\sum_{j(\neq 0)}R_j\Phi_{0j}',\ \ \ 
 v^s = \frac{1}{2\Omega}\sum_{j(\neq 0)}R_j\Phi_{0j}'s_j,
 \label{eq:vvs} \\
&& w = \frac{1}{2\Omega}\sum_{j(\neq 0)}R_j^2\Phi_{0j}'',\ \ \ 
 w^s = \frac{1}{2\Omega}\sum_{j(\neq 0)}R_j^2\Phi_{0j}''s_j,
 \label{eq:wws} 
\end{eqnarray}
with $s_j = (x_j^4 + y_j^4 + z_j^4)/R_j^4$.
Together with $\alpha_0$, three of other quantities representing
the strength of environment-dependence are defined:
\begin{eqnarray}
&&\alpha_0^s = \sum_{k(\neq 0)}R_{k}\rho'(R_k)s_k,
\label{eq:alpha-s} \\
&&\beta_0 =  \sum_{k(\neq 0)}R_{k}^2\rho''(R_k),\ \ \ 
\beta_0^s =  \sum_{k(\neq 0)}R_{k}^2\rho''(R_k)s_k.
\label{eq:beta-s} 
\end{eqnarray}

\begin{figure}[tb]
\begin{center}
\includegraphics[height=50mm, trim=0 15 0 0]{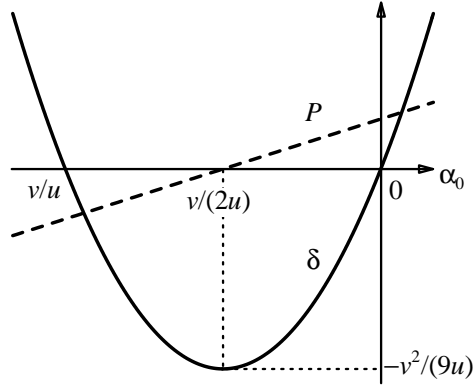}
\end{center}
\caption{The Cauchy violation, $\delta$, and pressure, $P$, as a function 
of strength of environment-dependence, $\alpha_0$, 
for the repulsive potential. }
\label{P-delta-vs-alpha0}
\end{figure}

Equations (\ref{eq:P}) and (\ref{eq:delta}),  seen as a
linear and quadratic functions of $\alpha_0$ respectively,
explain how the negative Cauchy violation occurs when
an environment-dependence is introduced, as presented
instructively in  Fig.~\ref{P-delta-vs-alpha0}.
It is to be noted that $\alpha_0=v/(2u)$ is unphysical point where
our `repulsive' potential is found no longer repulsive, giving
$P=0$.
Therefore, the magnitude of dimensionless parameter 
$\alpha_0$ should usually be much smaller than $|v/(2u)|$.
An instructive example may be the case of inverse-power-law potential, 
$\Phi\propto R^{-n}$, in which the critical value can be easily 
found to be $v/(2u)=-n/2$.
These analysis should be useful in understanding
how the environmental dependence in repulsive worked for
the problem of small or negative Cauchy 
pressure
 ($C_{12}-C_{44}$ for cubic systems, 
 $C_{13}-C_{44}$ and $C_{12}-C_{66}$ 
for tetragonal or hexagonal systems)
in transition metals and intermetallic 
compounds\cite{DucPZV88,MrovecNPV04}.
In these covalently bonded systems at equilibrium, 
a negative pressure $P_\mathrm{bond}$ from 
the attractive covalent bond energy 
counterbalances the positive one from
the repulsion. In Fig.~\ref{P-delta-vs-alpha0},
this situation corresponds to the `high pressure' case 
in which $P=|P_\mathrm{bond}|$ and a negative contribution of $\delta$
with $v/(2u)<\alpha_0<0$.

Finally, parameterisations for  functions
$V_{R}(R)$ in Eq.(\ref{eq:separable-Phi})
and $\rho(R)$ in Eq.(\ref{eq:lanmda_i}) must be determined.
They are basically similar and described by a superposition of 
the square of two-center overlap integrals.
Using Eqs.~(\ref{Phi(R)-from}) and (\ref{kappa-eqn}) as a guide, 
we employ the following functions, namely
\begin{equation}
V_{R}(R) = A(1+a_{1}R+a_{2}R^2)
\exp({-\mu_{1}R-\mu_{2}R^2})
\end{equation}
and
\begin{equation}
\rho(R) = g\exp({-\nu{}R}),
\end{equation}
where the parameters $A,\mu_1,g,\nu$ are essential, 
and $a_1, a_2, \mu_2$ are for flexibility
of the fitting.

We do not explicitly include the pairwise $E_{\mathrm{rep}}$ in
the present model, because it is actually unknown, but may not be dominant, 
and therefore we may think of it as being absorbed effectively 
in the pairwise component of overlap repulsion unless it proves significant.

\section{Results}

\begin{table}
\caption{Fitted parameters in $V_R$ and $\rho$.} 
\begin{indented}
\lineup
\item[]\begin{tabular}{@{}*{7}{l}}
\br                              
\0\0\0$A$ (J)&$\m a_1$ (\AA$^{-1}$)&$a_2$ (\AA$^{-2}$)&
$\mu_1$ (\AA$^{-1}$)&\m$\mu_2$ (\AA$^{-2}$)&\0$g$&$\0\nu$ (\AA$^{-1}$)\cr
\mr
\02.10$\times10^{-15}$ & \0$-0.5819$ & 0.09309 &
 \03.000 & $-0.03996$ & 80.0 & \03.60\cr 
\br
\label{table1}
\end{tabular}
\end{indented}
\end{table}

\begin{figure}[tb]
\begin{center}
\includegraphics[height=68mm, trim=0 25 0 0]{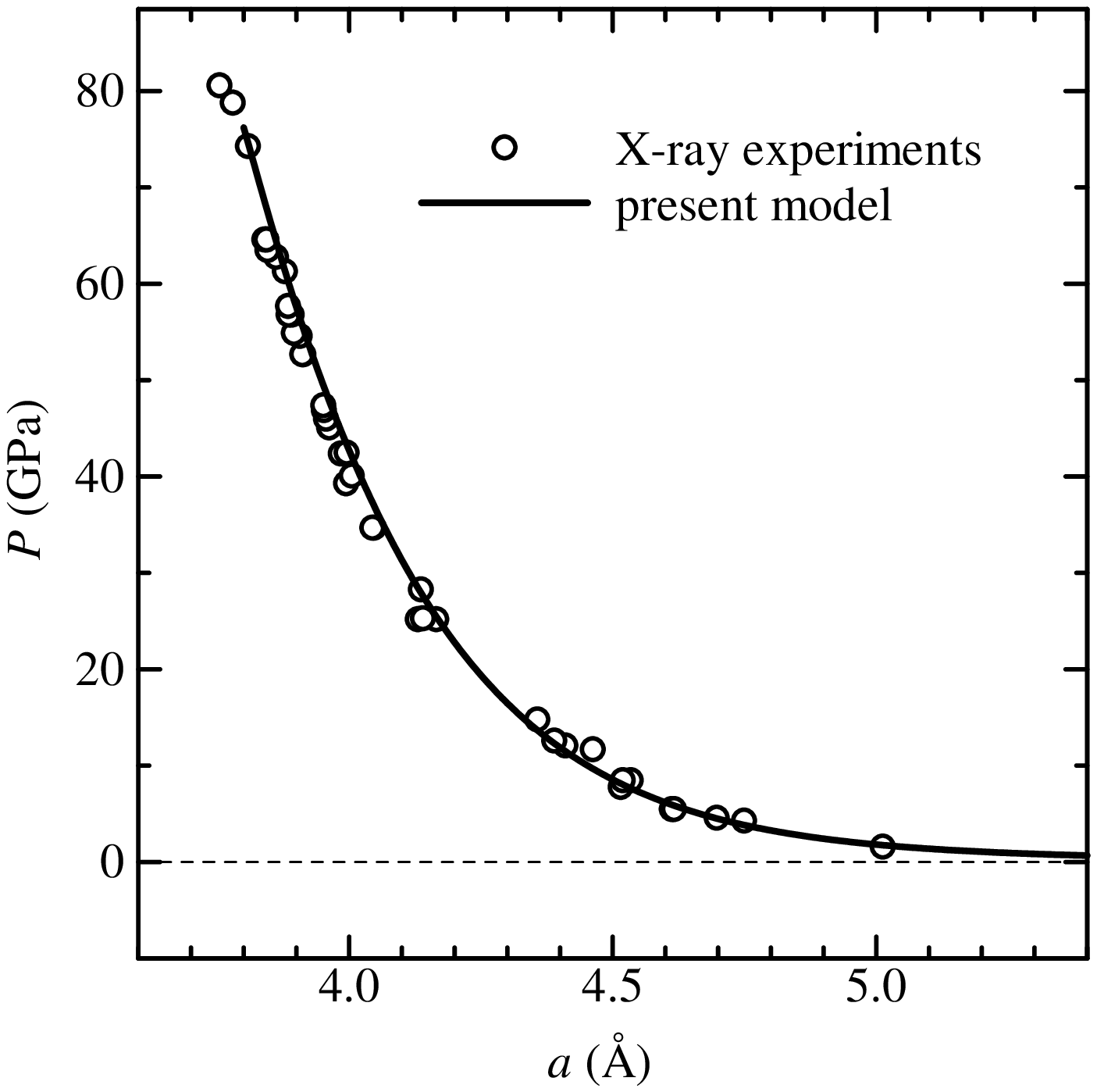}
\hspace{3mm}
\includegraphics[height=70mm, trim=0 15 0 0]{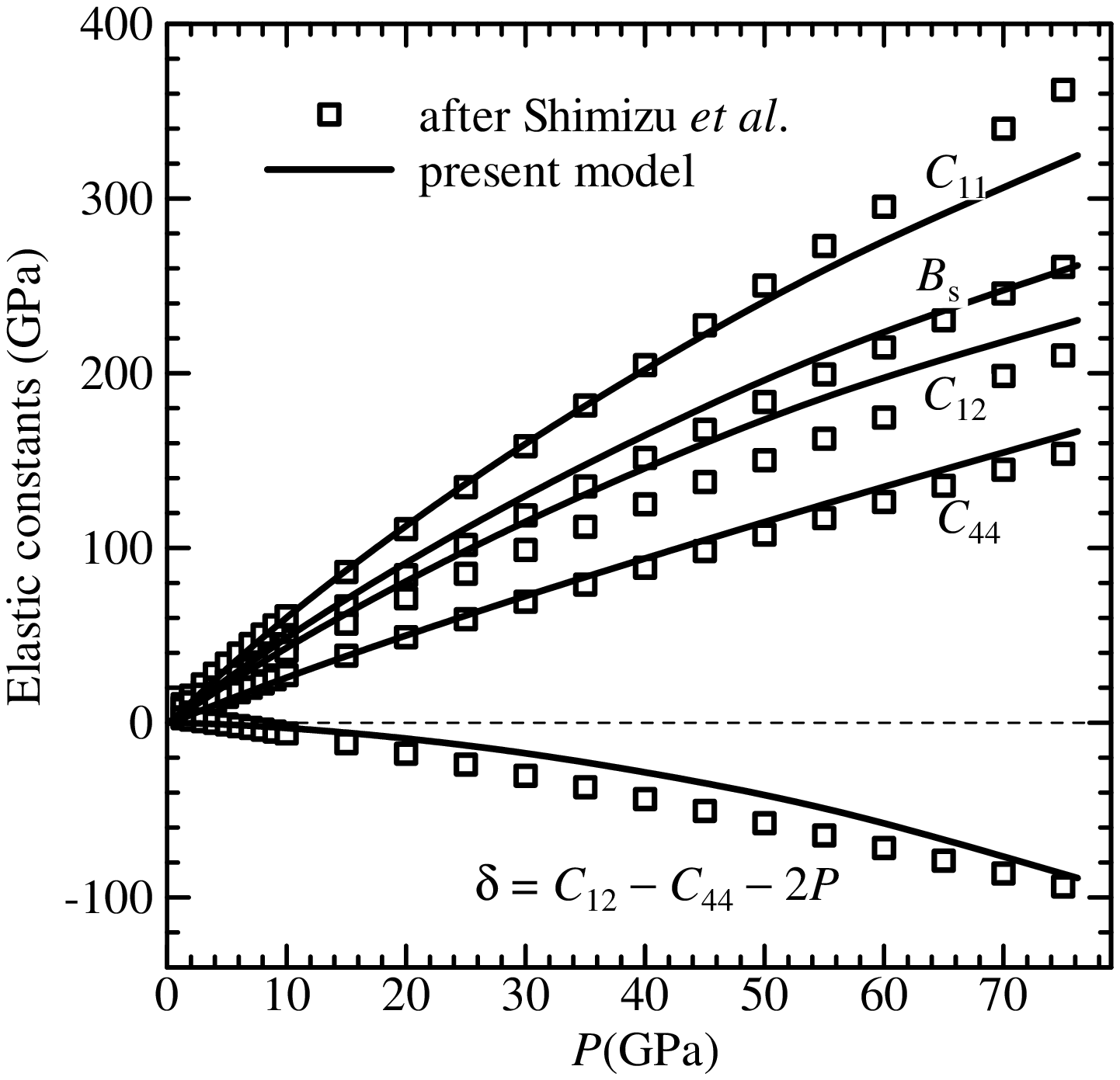}
\end{center}
\caption{Left: Pressure versus lattice constant. 
Circles denote X-ray observation. 
(see references in \cite{ShimizuTKS01}) 
Right: Elastic constants versus pressure: 
The present model (solid lines) 
compared with experimental results by Shimizu 
{\it et al.} \cite{ShimizuTKS01}.}
\label{fig:3}
\end{figure}

Using the model of overlap repulsion plus
the pairwise van der Waals potential ($-c_6R_{ij}^{-6}$) with
the Lennard-Jones parameters for argon\cite{Kittel05},
{\it i.e.} $c_6=4\varepsilon\sigma^6$ with  
$\varepsilon=1.67\times10^{-21}$ J and $\sigma = 3.40$ \AA,
the parameters are fitted to the results\cite{Aoki03} by 
{\it ab initio} full-potential LMTO calculations
with the generalized gradient approximation of PW91
 (GGA-PW91)\cite{GGA-PW91}
for the exchange-correlation,
since GGA-PW91 results are remarkably in good agreement 
with the experimental results for solid argon. 
However, GGA-PW91 is known to predict always positive pressures,
and a small positive pressure even at the experimental lattice 
constant $a=5.13$ \AA for the equilibrium at zero pressure. 
The adjusted parameters are listed in Table~\ref{table1}.

Figure~\ref{fig:3} shows the elastic properties
of fcc solid argon predicted by the present model compared
with the experimental results. The agreement is impressive.
However, a negative curvature in predicted $\delta$ 
at very high pressures can be seen as a small deviation, which is
reflected in $B$, $C_{11}$, $C_{12}$ (and not in $C_{44}$)
as it can be understood from $\delta$-term (and its absence) 
in Eqs.(\ref{eq:Bs})-(\ref{eq:C44PKd}). This will be corrected
if we design more flexible function for $\rho$. But if we do so,
we also need to evaluate neglected terms, for example, the higher order
many-atom effects due to $(1+\mathsf{O})^{-1}$ factor, which
will be handled in a separate study.

The stability problem of fcc against hcp is still very subtle
even with the present environment-dependent model without 
zero point energy. The result is very sensitive to the cutoff.
For example, using only the present repulsive model (without $E_\mathrm{vdW}$), 
the fcc--hcp difference in enthalpy is evaluated to be
$-0.01$ meV at 20 GPa and  $0.11$ meV at 60 GPa if we include 86 neighbours
within 6 shells in fcc and equivalently within 9 shells in hcp. 
The difference in the zero point energy\cite{Rosciszewski-etal00} 
would be dominant. 
Therefore
the environmental or many-atom effect in overlap repulsion
may not be a remedy for the problem of fcc stability.

\section{Conclusion}

We have developed an analytic and environment-dependent
interatomic potential for the overlap repulsion in
solid argon. The functional form, 
of environment-dependence in particular, is simple and 
physically transparent,
being based on the non-orthogonal Tight-Binding theory for the 
closed-shell systems. 
The present model was shown to well reproduce the observed 
elastic properties of solid argon including
the Cauchy violation at high pressures.

A useful and novel analysis has clearly demonstrated 
how the elastic properties 
are related to the environment-dependence incorporated
into a generic pairwise potential.
It is speculated that the present functional provides
not only excellent description for elastic properties
of a solid noble gas, but also useful description for
the problem of small or negative Cauchy pressures 
in covalently bonded systems.

\section{Acknowledgments}
The authors would like to thank Dr. Duc Nguyen-Manh for useful
conversations about their works on Cauchy pressure,
Prof. Y. Shimizu and S. Sasaki for helpful information of their
experiments, and Dr. T. Iitaka for stimulating information at
the onset of the present study. 
MA thanks the Kogyo-Club of the Faculty of
Engineering, Gifu University for financial support.


\section*{References}

\begin{thebibliography}{99}
%
\bibitem{ShimizuTKS01} Shimizu, H., Tashiro, H., Kume, T. and 
Sasaki, S.,  Phys. Rev. Lett. {\bf 86}, 4568 (2001).
%
\bibitem{Grimsditch-etal86} Grimsditch, M., Loubeyre, P. and Polian, A., 
 Phys. Rev. {\bf B 33}, 7192 (1986).
%
\bibitem{BornHuang88} Born, M. and Huang, K., 
 {\it Dynamical theory of crystal lattices}, 
 (Oxford University Press, 1954).
%
\bibitem{Aziz-Slaman90} Aziz, R. A. and  Slaman, M. J., 
J. Chem. Phys. {\bf 92}, 1030 (1990).
%
\bibitem{Iitaka-E-02} Iitaka, T. and Ebisuzaki, T., 
Phys. Rev. {\bf B 65}, 012103 (2002).
%
\bibitem{Tse-etal-02} Tse, J.S. Klug, D.D., Shpakov, V, and 
Rodgers, J.R., Solid State Commun., {\bf 122}, 575 (2002).
%
\bibitem{Tsuchiya-K-02} Tsuchiya, T. and Kawamura, K., 
J. Chem. Phys. {\bf 117}, 5859 (2002), and references therein.
%
\bibitem{Aoki03} Aoki, M., Rev. High Pressure Sci. Techn. {\bf 13} 
(in Japanese), 218 (2003).
%
\bibitem{Rosciszewski-etal00} Rosciszewski, K., Paulus, B., Flude, P. 
 and Stoll, H., 
 Phys. Rev. {\bf B 62}, 5482 (2000).
%
\bibitem{Lotrich-S97} Lotrich, V.F. and Szalewicz, K., 
 Phys. Rev. Lett. {\bf 79}, 1301 (1997).
%
\bibitem{Schwerdtfeger-etal06} Schwerdtfeger, P., Gaston, N.,
Krawczyk, R.P., Tonner, R. and Moyano, G.E., 
 Phys. Rev. {\bf B 73}, 064112 (2006).
%
\bibitem{SavrasovS92} Savrasov, S.Yu. and Savrasov, D.Yu., 
 Phys. Rev. {\bf B 46}, 12181 (1992).
%
\bibitem{TangWCH96} Tang, M.S., Wang, C.Z., Chan, C.T. and Ho, K.M., 
Phys. Rev. {\bf B 53}, 979 (1996).
%
\bibitem{HaasWFEH98} Haas, H., Wang, C.Z., Fahnle, M., Elsasser, C. 
and Ho, K.M., Phys. Rev. {\bf B 57}, 1461 (1998).
%
\bibitem{MrovecNPV04} Mrovec, M., Nguyen-Manh, D., 
Pettifor, D.G. and Vitek, V., Phys. Rev. {\bf B 69}, 94115 (2004).
%
\bibitem{DucPZV88} Nguyen-Manh, D., Pettifor, D.G., 
 Znum, S. and Vitek, V.,  {\it Tight-Binding approach 
 to computational Materials Science} , Turchi, P.E.A., 
Gonis, A., and Colombo, L., (Eds.)
(Mater. Res. Soc. Symp. Proc. 491), pp 353-358, (MRS, 1998).
%
\bibitem{Finnis03} M.W. Finnis, 
 {\it Interatomic forces in condensed matter},
 (Oxford University Press, 2003).
%
\bibitem{SFPO88} 
 Sutton, A.P., Finnis, M.W., Pettifor D.G. and Ohta, Y.,
 J. Phys. C {\bf 21}, 35 (1988). 
%
\bibitem{Pettifor95} Pettifor, D.G., 
{\it Bonding and structure in molecules and solids}
 (Oxford University Press, 1995).
%
\bibitem{DucPV00} Nguyen-Manh, D., Pettifor and Vitek, V., 
 Phys. Rev. Lett. {\bf 85}, 4136 (2000).
%
\bibitem{Skinner-P-91} Skinner, A.J. and Pettifor, D.G.,
J. Phys.:Cond. Mat. {\bf 3}, 2029 (11991).
%
\bibitem{Harris85} Harris, J., 
Phys. Rev., {\bf B 31}, 1770 (1985).
%
\bibitem{Foulkes89} Foulkes, W.M.C. and Haydock, R., 
Phys. Rev., {\bf B 39}, 12520 (1989).
%
\bibitem{Sangster73} Sangster, M.J.L., 
J. Phys. Chem. Solids. {\bf 34}, 355 (1973).
%
\bibitem{Matsui05} Matsui, M., 
J. Chem. Phys. {\bf 108}, 3304 (1998).
%
\bibitem{Marks-FHP-01} 
Marks, N.~A., Finnis, M.~W., Harding, J.~H. and Pyper, N.~C.
J. Chem. Phys. {\bf 114}, 4406 (2001).
%
\bibitem{Kittel05}  Kittel, C., 
{\it Introduction to solid state physics} Eighth edition. 
(John Wiley \& Sons, 2005)
%
\bibitem{GGA-PW91} Perdew, J.~P., in Electronic Structure of Solids'91,
eds. Ziesche, P., Eschrig,  H. (Akademie Verlag, Berlin, 1991).

\end{thebibliography}
\end{document}